\documentclass[prl,twocolumn,amsmath,amssymb,floatfix]{revtex4-1}
\usepackage{graphicx}% Include figure files
\usepackage{dcolumn}% Align table columns on decimal point
\newcolumntype{d}[1]{D{.}{.}{#1}}
\usepackage{bm}% bold math
\usepackage{longtable}
\usepackage{mathtools}

\usepackage{epstopdf}
\usepackage{float}

\begin{document}

\title{Trigonal-to-monoclinic structural transition in TiSe$_2$ due
  to a combined condensation of $\mathbf{ \textit{q} = (\frac{1}{2},0,0)}$ and
  $\mathbf{(\frac{1}{2},0,\frac{1}{2})}$ phonon instabilities}

\author{Alaska Subedi} 

\affiliation{CPHT, CNRS, Ecole Polytechnique, IP Paris, F-91128
  Palaiseau, France} 
% \affiliation{Coll\`ege de France, 11 place
%  Marcelin Berthelot, 75005 Paris, France}

\date{\today}

\begin{abstract}

I present first principles calculations of the phonon dispersions of
TiSe$_2$ in the $P\overline{3}c1$ phase, which is the currently
accepted low-temperature structure of this material.  They show weak
instabilities in the acoustic branches in the out-of-plane direction,
suggesting that this phase may not be the true ground state.  To find
the lowest energy structure, I study the energetics of all possible
distorted structures corresponding to the isotropy subgroups of
$P\overline{3}m1$ for the $M_1^-$ and $L_1^-$ phonon instabilities
present in this high-temperature phase at $q = (\frac{1}{2},0,0)$ and
$(\frac{1}{2},0,\frac{1}{2})$, respectively.  I was able to stabilize
10 different structures that are lower in energy relative to the
parent $P\overline{3}m1$ phase, including two monoclinic structures
more energetically stable than the $P\overline{3}c1$ phase.  The
lowest energy structure has the space group $C2$ with the order
parameter $M_1^- (a,0,0) + L_1^- (0,b,b)$.  This structure lacks
inversion symmetry, and its primitive unit cell has 12 atoms.

\end{abstract}

% \pacs{64.60.Ej,63.20.D-,71.30.+h}

\maketitle

\section{Introduction}

The structural transition near 200 K in $1T$-TiSe$_2$ has been
frequently studied since its three-directional superlattice was
reported by Di Salvo \textit{et al.}  in 1976 \cite{disa76}.  A phonon
softening at the wave vector $(\frac{1}{2},0,\frac{1}{2})$ in the
parent phase of this material has been unambiguously identified
\cite{holt01,weber11}, but the microscopic mechanism underlying this
charge density wave (CDW) transition is still being debated.  The
parent phase of TiSe$_2$ is either a semimetal or a semiconductor with
a low carrier concentration
\cite{disa76,stof85,ande85,pill00,kidd02,ross02,cui06,qian07,li07,cerc07,rasc08,rohw11,chen16,mott19},
which precludes an explanation based on Fermi surface nesting. Hence,
other mechanisms such as excitonic condensation
\cite{wils77,monn09,mohr11,may11,koga17}, Jahn-Teller effect
\cite{hugh77,whan92,wegn20}, incipient antiferroelectricity
\cite{whit77,buss02}, electron-phonon coupling
\cite{yosh80,moti81a,moti81b,suzu85}, or some combination thereof
\cite{weze10a,weze10b,monn11,pore14,kane18} has been invoked to
explain this transition.

The high-temperature phase of TiSe$_2$ occurs in a trigonal structure
with the space group $P\overline{3}m1$ \cite{ofte28,riek76}.  This
structure is composed of hexagonal layers of Ti sandwiched between two
hexagonal layers of Se such that the Ti ions are situated inside Se
octahedra.  Each layer has three twofold rotational axes and three
mirror planes along and perpendicular, respectively, to the three
chains forming the hexagonal lattice. The low-temperature phase has
been reported to form a $2\times2\times2$ superlattice with the space
group $P\overline{3}c1$ \cite{wils78}.  In this structure, all the
three twofold rotational symmetries present in each layer are broken.
However, the presence of a glide plane restores the twofold rotational
symmetries in the full lattice.

There are experimental indications that further rotational, mirror,
and inversion symmetries are broken in the low-temperature phase.
Ishioka \textit{et al.} have claimed that the CDW phase in this
material is chiral based on their scanning tunneling microscopy (STM)
experiments \cite{ishi10,ishi11}.  Such a chiral phase has been
theoretically understood as a form of orbital ordering
\cite{weze11,weze12,grad15}, and there are experimental evidences supporting
this claim \cite{ivar12,cast13,peng21}.
However, more recent STM experiments have questioned this
interpretation and suggest that the CDW phase is achiral
\cite{nove15,hild18}.  In the midst of this debate
\cite{lin19,rose19,ueda21}, Xu \textit{et al.} have reported the
measurements of circular photogalvanic effect current that suggests the
presence of a low-symmetry structure without inversion symmetry below
174 K \cite{xu20}.  But this gyrotropic phase has been argued to occur
only in the photoexcited state \cite{wick20}.

The electronic properties of TiSe$_2$ and the structural instability
of its high-temperature phase has been extensively studied using
density functional theory (DFT) based first principles calculations
\cite{chen16,zung78,fang97,jish08,cala11,cazza12,zhu12,vydr15,bian15,hell17,hell21}.
However, neither the structural stability of the $P\overline{3}c1$ CDW
phase nor a detailed study of all possible structures arising out of
the phonon instabilities present in the parent phase has been
investigated using DFT calculations.  In particular, the energetics of
the low-symmetry structures resulting from a combined condensation of
the phonon instabilies at $M$ $(\frac{1}{2},0,0)$ and $L$
$(\frac{1}{2},0,\frac{1}{2})$ has not been explored.  A theoretical
study examining these aspects would be helpful in answering whether a
structure with broken inversion symmetry is the true ground state of
pure TiSe$_2$ or it is induced by external stimuli such as defects and
photoexcitations.

In this paper, I present the calculated phonon dispersions of the
$2\times2\times2$ $P\overline{3}c1$ phase, which show acoustic
branches with weak instabilities in the out-of-plane direction.  This
suggests that the $P\overline{3}c1$ structure may not be the true
ground state of this material.  To find the lowest energy structure, I
generated all possible distortions corresponding to the isotropy
subgroups that can arise due to the phonon instabilities at the $M$
and $L$ points present in the parent $P\overline{3}m1$ phase of the
material.  After full structural relaxations minimizing both the
forces and stresses, I was able to stabilize 10 different structures
that are lower in energy than the parent $P\overline{3}m1$ phase.
These include two monoclinic structures that are more energetically
stable than the $P\overline{3}c1$ phase.  The lowest energy structure
has the space group $C2$ with the order parameter $M_1^- (a,0,0) +
L_1^- (0,b,b)$. This structure has no inversion symmetry, and its
primitive unit cell has 12 atoms.

\section{Computational Approach}

The phonon dispersions and structural relaxation calculations
presented here were performed using the pseudopotential-based {\sc
  quantum espresso} package \cite{qe}.  I used the pseudopotentials
generated by Dal Corso \cite{dalc14} and energy cutoffs of 60 and 600
Ry for the basis-set and charge density expansions, respectively.  The
calculations were performed using the optB88-vdW exchange-correlation
functional that accurately treats the van der Waals interaction
\cite{optb88}.  In the phonon calculations, $24\times24\times12$ and
$12\times12\times6$ $k$-point grids were used for the Brillouin zone
integration in the $P\overline{3}m1$ and $P\overline{3}c1$ phases,
respectively. Dynamical matrices were calculated on a
$8\times8\times4$ grid for the $P\overline{3}m1$ phase and
$4\times4\times4$ grid for the $P\overline{3}c1$ phase using density
functional perturbation theory \cite{dfpt}, and Fourier interpolation
was used to obtain the phonon dispersions.  I used the {\sc isotropy}
package to enumerate all the order parameters that are possible due to
the unstable phonon modes $M_1^-$ and $L_1^-$ of the parent phase
\cite{isotr}. Structural relaxation calculations of the structures
corresponding to different isotropy subgroups were performed on
$2\times2\times2$ supercells using a $20\times20\times10$ $k$-point
grid.  I checked the relative energy orderings of the two lowest
energy structures using a $24\times24\times12$ $k$-point grid and 85
Ry basis-set cutoff.  A 0.01 Ry Marzari-Vanderbilt smearing was used
in all the calculations.

I made extensive use of the {\sc findsym} \cite{findsym}, {\sc
  amplimodes} \cite{ampli}, {\sc spglib} \cite{spglib}, and {\sc
  phonopy} \cite{phonopy} packages in the symmetry analysis of the
relaxed structures.  A previous study has shown that the spin-orbit
interaction does not modify the structural instability
of this material \cite{hell17}, so it was neglected in all the
calculations presented in this paper.

\section{Results and Discussion}

\begin{figure}
  \includegraphics[width=\columnwidth]{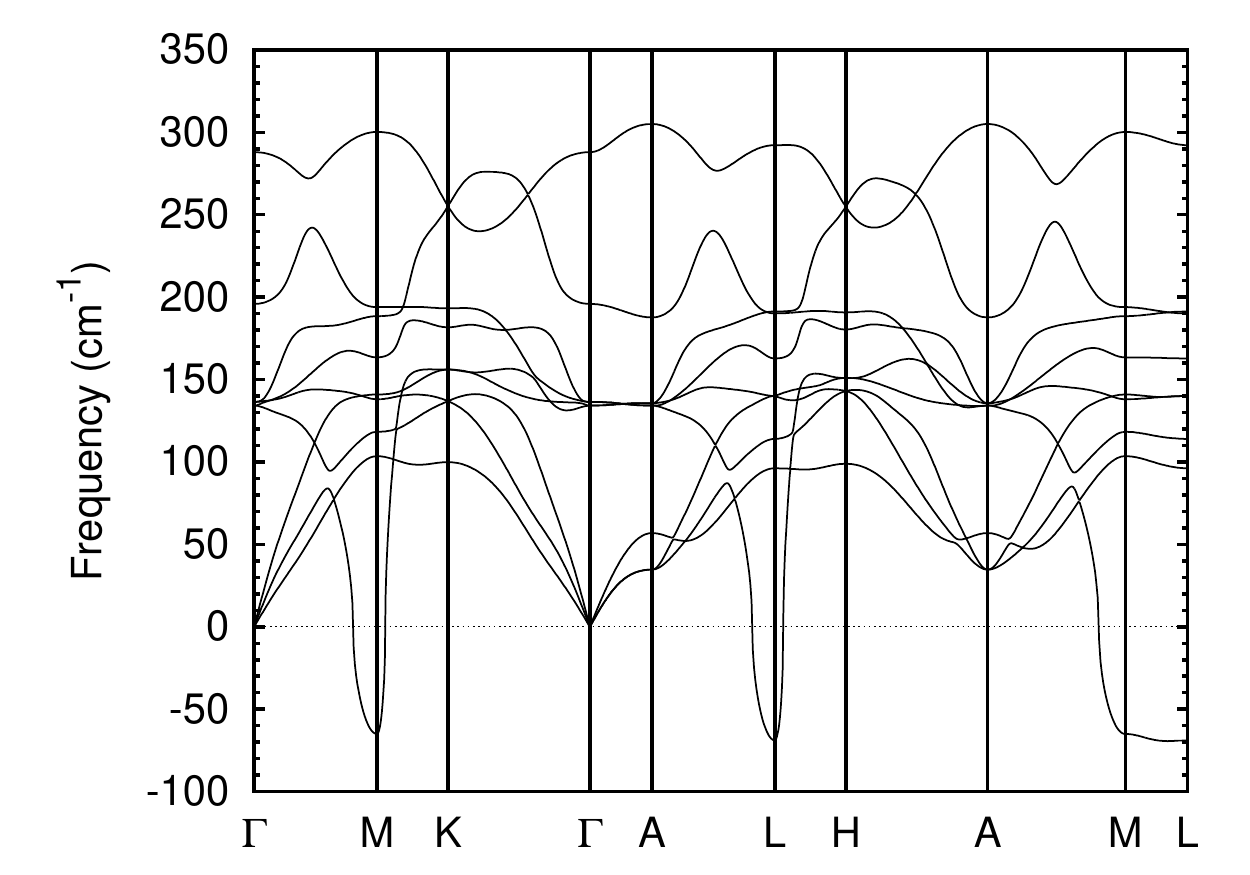}
  \caption{Calculated phonon dispersions of TiSe$_2$ in the
    parent $P\overline{3}m1$ phase calculated using the
    optB88-vdW functional.  The high-symmetry points are $\Gamma$
    $(0,0,0)$, $M$ $(\frac{1}{2},0,0)$, $K$
    $(\frac{1}{3},\frac{1}{3},0)$, $A$ $(0,0,\frac{1}{2})$, $L$
    $(\frac{1}{2},0,\frac{1}{2})$, and $H$
    $(\frac{1}{3},\frac{1}{3},\frac{1}{2})$ in terms of the reciprocal
    lattice vectors.}
  \label{fig:phhT}
\end{figure}

The calculated optB88-vdW phonon dispersions of the fully-relaxed
TiSe$_2$ in the parent $P\overline{3}m1$ structure is shown in
Fig.~\ref{fig:phhT}.  They agree well with the previous calculations
\cite{cala11,hell17}.  The calculated values of the $A_g$ 196
cm$^{-1}$ and highest-frequency $E_u$ 135 cm$^{-1}$ modes also compare
well with the experimental values of $A_g$ 200 cm$^{-1}$ \cite{suga80}
and $E_u$ 137 cm$^{-1}$ \cite{holy77}.
There is a phonon branch that is unstable along the path $M$--$L$.
Both $M$ $\left\{\left(0,\frac{1}{2},0\right),
\left(\frac{1}{2},0,0\right), \left(\frac{1}{2},\frac{1}{2},0\right)
\right\}$ and $L$ $\left\{\left(0,\frac{1}{2},\frac{1}{2}\right),
\left(\frac{1}{2},0,\frac{1}{2}\right),
\left(\frac{1}{2},\frac{1}{2},\frac{1}{2}\right) \right\}$ have three
elements in their star.  Hence, even though the unstable branch is
nondegenerate, several low-symmetry structures are possible due to
these instabilities.  The instability at $L$ is slightly stronger than at
$M$, and the low-temperature CDW phase of this material has been
understood to form due to the simultaneous condensation at the three
wave vectors belonging to $L$ \cite{disa76}.  Indeed, Bianco
\textit{et al.} have performed a detailed DFT-based theoretical study
and found that the energy gain due to the triple-$q$ condensation at
$L$ is larger than the triple-$q$ condensation at $M$ as well as
single-$q$ condensations at $L$ and $M$ \cite{bian15}.

\begin{figure}
  \includegraphics[width=\columnwidth]{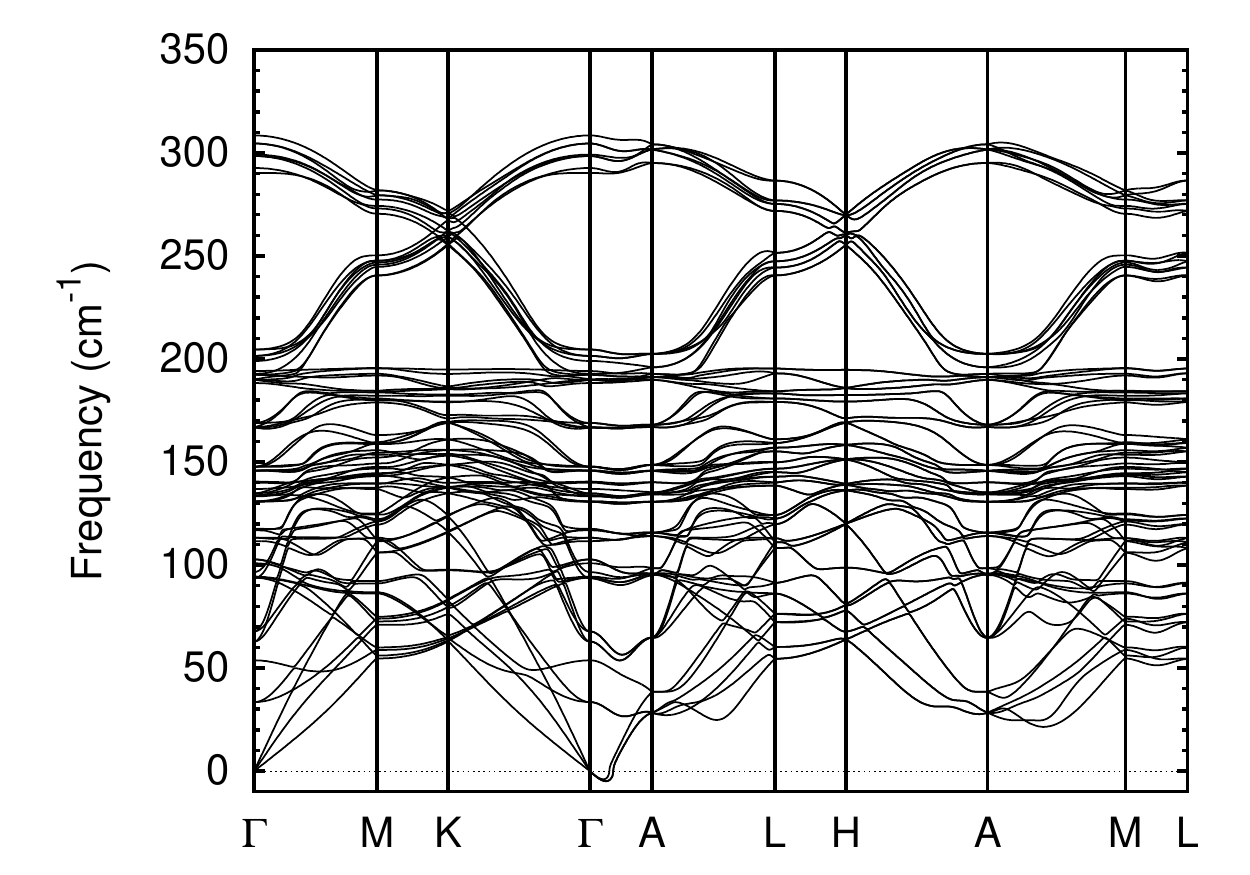}
  \caption{Calculated phonon dispersions of TiSe$_2$ in the $L_1^-
    (a,a,a)$ $P\overline{3}c1$ phase calculated using the optB88-vdW
    functional. The acoustic branches are unstable in the out-of-plane
    direction $\Gamma$--$A$.}
  \label{fig:phlT}
\end{figure}

Although the structural instability of the high-temperature phase of
TiSe$_2$ has been extensively studied using DFT-based calculations
\cite{weber11,cala11,bian15,hell17}, the relative energetic stability
of all possible structures arising due to the instabilities at $M$ and
$L$ has yet to be investigated.  In fact, the structural stability of
the currently accepted low-temperature triple-$q$ $P\overline{3}c1$
phase has not been confirmed theoretically despite there being
experimental evidences that the low-temperature structure has a
symmetry lower than trigonal \cite{ishi10,xu20}.  I calculated the
phonon dispersions of the fully-relaxed $P\overline{3}c1$ phase, which
is shown in Fig.~\ref{fig:phlT}.  I find that all the optical phonon
branches are stable.  However, the acoustic branches show weak
instabilities in the out-of-plane $(0,0,q_z)$ direction. The
instabilities occur for $q_z < \frac{1}{6}$, which is not in the
$4\times4\times4$ grid used to calculate the dynamical matrices.  To
confirm the presence of the instabilites, I calculated the dynamical
matrices at $q_z = \frac{1}{16}$ and $\frac{1}{24}$, which yielded
three modes with imaginary frequencies. This suggests that the
currently accepted low-temperature $P\overline{3}c1$ structure may not
be the true ground state of this material.

\begin{table}
    \caption{\label{tab:iso} Isotropy subgroups of $P\overline{3}m1$
      for the representations $L_1^-$ and $M_1^-$, and the
      corresponding six-dimensional order parameters in the subspace
      spanned by the stars of $M$
      $\left\{\left(0,\frac{1}{2},0\right),
      \left(\frac{1}{2},0,0\right),
      \left(\frac{1}{2},\frac{1}{2},0\right) \right\}$ and $L$
      $\left\{\left(0,\frac{1}{2},\frac{1}{2}\right),
      \left(\frac{1}{2},0,\frac{1}{2}\right),
      \left(\frac{1}{2},\frac{1}{2},\frac{1}{2}\right)
      \right\}$. Total energies of the structures corresponding to
      these order parameters after full structural relaxations
      minimizing the atomic forces and lattice stresses are given in
      the units of meV per formula unit relative to the parent
      $P\overline{3}m1$ phase.  Not all distortions could be
      stabilized.}
    
    \begin{ruledtabular}  
      %       \begin{tabular}{l c c d{3.3}}
      \begin{tabular}{l c c c}
      space group (\#num.) & $M_1^-$ & $L_1^-$ &  energy (meV/f.u.) \\ % \multicolumn{1}{c}{energy (meV/f.u.)}\\
      \hline
      $P\overline{3}m1$ (\#164)  & $(0,0,0)$ & $(0,0,0)$ & $\phantom{-}0.000$  \\
      $P2/c$          (\#13)     & $(a,0,0)$ & $(0,0,0)$ & $-0.726$ \\
      $C2/c$ (\#15)              & $(0,0,0)$ & $(a,0,0)$ & $-0.755$ \\
      $C2/m$          (\#12)     & $(a,a,0)$ & $(0,0,0)$ & $-1.004$ \\
      $P\overline{1}$ (\#2)      & $(a,0,0)$ & $(0,b,0)$ & $-1.031$ \\
      $C2/m$ (\#12)              & $(0,0,0)$ & $(a,a,0)$ & $-1.046$ \\
      $P321$          (\#150)    & $(a,a,a)$ & $(0,0,0)$ & $-1.136$ \\
      $C2/c$ (\#15)              & $(a,a,0)$ & $(0,0,b)$ & $-1.170$ \\
      $P\overline{3}c1$ (\#165)  & $(0,0,0)$ & $(a,a,a)$ & $-1.184$ \\
      $C2/c$ (\#15)              & $(0,0,0)$ & $(a,a,b)$ & $-1.188$ \\
      $C2$            (\#5)      & $(a,0,0)$ & $(0,b,b)$ & $-1.192$ \\
      %
%      $P2/c$          (\#13)     & $(a,0,0)$ & $(0,0,0)$ & $-0.726$ \\
%      $C2/m$          (\#12)     & $(a,a,0)$ & $(0,0,0)$ & $-1.004$ \\
%      $P321$          (\#150)    & $(a,a,a)$ & $(0,0,0)$ & $-1.136$ \\
      $P\overline{1}$ (\#2)      & $(a,b,0)$ & $(0,0,0)$ & --- \\
      $C2$            (\#5)      & $(a,a,b)$ & $(0,0,0)$ & --- \\
      $P1$            (\#1)      & $(a,b,c)$ & $(0,0,0)$ & --- \\
      %
%      $C2/c$ (\#15)              & $(0,0,0)$ & $(a,0,0)$ & $-0.755$ \\
%      $C2/m$ (\#12)              & $(0,0,0)$ & $(a,a,0)$ & $-1.046$ \\
%      $P\overline{3}c1$ (\#165)  & $(0,0,0)$ & $(a,a,a)$ & $-1.184$ \\
      $P\overline{1}$ (\#2)      & $(0,0,0)$ & $(a,b,0)$ & --- \\
%      $C2/c$ (\#15)              & $(0,0,0)$ & $(a,a,b)$ & $-1.188$ \\
      $P\overline{1}$ (\#2)      & $(0,0,0)$ & $(a,b,c)$ & --- \\  % -1.188 \\
      $P2/c$          (\#13)     & $(a,0,0)$ & $(b,0,0)$ & --- \\
%      $P\overline{1}$ (\#2)      & $(a,0,0)$ & $(0,b,0)$ & $-1.031$ \\
%      $C2/c$ (\#15)              & $(a,a,0)$ & $(0,0,b)$ & $-1.170$ \\
      $P\overline{1}$ (\#2)      & $(a,b,0)$ & $(0,0,c)$ & --- \\
%      $C2$            (\#5)      & $(a,0,0)$ & $(0,b,b)$ & $-1.192$ \\
      $C2/m$          (\#12)     & $(a,a,0)$ & $(b,b,0)$ & --- \\
      $C2/c$ (\#15)              & $(a,a,0)$ & $(b,-b,0)$ & --- \\
      $C2$            (\#5)      & $(a,a,b)$ & $(c,-c,0)$ & --- \\
      $P1$            (\#1)      & $(a,0,0)$ & $(0,b,c)$ & --- \\
      $P\overline{1}$ (\#2)      & $(a,b,0)$ & $(c,d,0)$ & --- \\
      $P321$          (\#150)    & $(a,a,a)$ & $(b,b,b)$ & --- \\
      $Cc$            (\#9)      & $(a,a,0)$ & $(b,-b,-c)$ & --- \\
      $C2$            (\#5)      & $(a,a,b)$ & $(c,c,d)$ & ---\\
      $P1$            (\#1)      & $(a,b,c)$ & $(d,e,f)$ & --- \\      
      \end{tabular}
    \end{ruledtabular}
\end{table}

The unstable phonon branch in the parent $P\overline{3}m1$ phase has
the representations $M_1^-$ and $L_1^-$ at $M$ and $L$, respectively.
I used the {\sc isotropy} package to determine all the isotropy
subgroups and order parameters that are possible due to these two
unstable phonons, which are listed in Table~\ref{tab:iso}.  I then
used the calculated phonon displacement vectors of the unstable modes
to generate all 26 possible distortions corresponding to the isotropy
subroups on $2\times2\times2$ supercells of the high-temperature
parent phase and fully relaxed these structures by minimizing both the
atomic forces and lattice stresses.

I was able to stabilize 10 different structures characterized by
distinct order parameters that have their calculated energies lower
than that of the high-temperature $P\overline{3}m1$ phase.  These
include the single- and triple-$q$ structures due to the $M_1^-$ and
$L_1^-$ instabilities discussed previously by Bianco \textit{et
  al.}\ \cite{bian15}.  Interestingly, there are three distinct
structures belonging to the same isotropy subrgoup $C2/c$ and two
structures with the subgroup $C2/m$.  The calculated total energies of
all these structures are given in Table~\ref{tab:iso}.  The energy
gain due to structural distortions are small, consistent with previous
results \cite{bian15}.  The $P\overline{3}c1$ structure is only $-1.184$
meV per formula unit (meV/f.u.) lower than the parent
$P\overline{3}m1$ phase. I find two more structures lower in energy
than the $P\overline{3}c1$ structure. They have space groups $C2/c$
and $C2$ with energies $-1.188$ and $-1.192$ meV/f.u.\ relative to the
parent phase, respectively.

\begin{figure}[t]
  \includegraphics[width=\columnwidth]{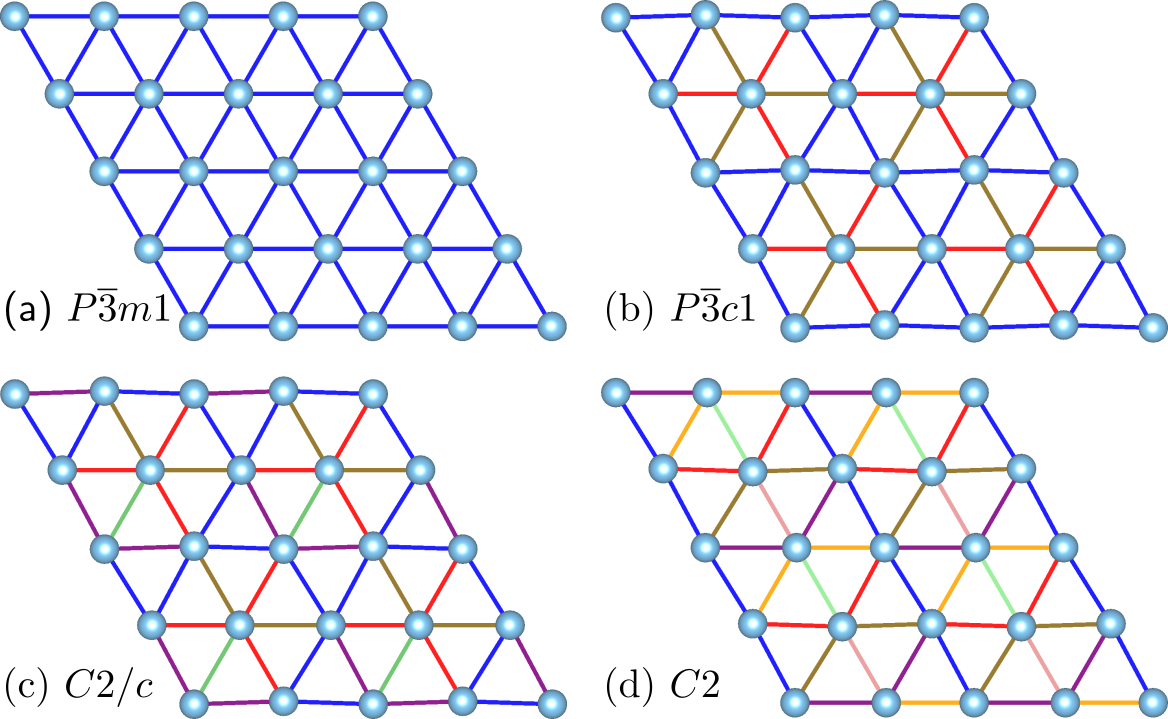}
  \caption{Ti hexagonal layers present in the (a) parent
    $P\overline{3}m1$, (b) $L_1^- (a,a,a)$ $P\overline{3}c1$, (c)
    $L_1^- (a,a,b)$ $C2/c$, and (d) $M_1^- (a,0,0) + L_1^- (0,b,b)$
    $C2$ phases of TiSe$_2$. There are one, three, five and seven
    nonequivalent Ti-Ti distances in the four phases, respectively,
    which are indicated by different colors. }
  \label{fig:layers}
\end{figure}

Fig.~\ref{fig:layers} shows the hexagonal Ti layer in the parent
$P\overline{3}m1$ and the three lowest energy structures with space
groups $P\overline{3}c1$, $C2/c$ and $C2$. Their full structural
parameters are given in the Supplemental Information \cite{supp}. In
the $P\overline{3}m1$ phase, all the Ti-Ti distances in the Ti
triangles are equal, and the calculated value of 3.5548 \AA\ is in
good agreement with experimentally determined one of 3.540
\AA\ \cite{riek76}.  Each element of the unstable mode at both $M$ and
$L$ causes nearest-neighbor antiparallel slidings within one set of the
three intersecting Ti chains that form the hexagonal lattice
\cite{disa76,bian15}.  This breaks the twofold rotational symmetries
the lie along the two other sets of Ti chains. The $P\overline{3}c1$
phase has the order parameter $L_1^- (a,a,a)$ and involves
simultaneous condensation of the unstable mode at all three wave
vectors in the star of $L$ with equal magnitudes. There are three
nonequivalent Ti-Ti distances in this phase.  The smallest calculated
Ti-Ti distance is 0.068 \AA\ shorter than the one in the parent phase,
which is in a reasonable agreement with the experimental value of 0.08
\AA\ \cite{disa76}.  Although all the twofold rotational symmetries
are broken within the hexagonal layers in this phase, the presence of
a $c$ glide plane restores the broken symmetries in the full
three-dimensional lattice.

The $C2/c$ phase that is lower in energy than the $P\overline{3}c1$
phase has the order parameter $L_1^- (a,a,b)$.  Since a component of
the order parameter is different along one direction, two additional
Ti-Ti distances become nonequivalent, for a total of five different
bond lengths in the hexagonal layer. This additionally breaks the
threefold rotational axis perpendicular to the hexagonal plane.
However, changes in the Ti-Ti distances due to this monoclinic
distortion is less than $2.0\times10^{-4}$ \AA\ relative to the
$P\overline{3}c1$ phase, and the monoclinic angle $\beta$ deviates
from $90^\circ$ by only $0.0016^\circ$.
% , reflecting the weak
% instabilities of the acoustic branches in the $P\overline{3}c1$ phase.

The lowest energy $C2$ phase involves condensation of both $M_1^-$ and
$L_1^-$ instabilities and has the order parameter $M_1^- (a,0,0) +
L_1^- (0,b,b)$.  Two more Ti-Ti distances become nonequivalent, and
this phase lacks the mirror as well as inversion symmetries present in
the $C2/c$ phase. The changes in the Ti-Ti distances in this structure
are up to $1.1\times10^{-3}$ \AA\ relative to the $P\overline{3}c1$
phase, which is larger than that calculated for the $C2/c$
structure. Unlike the $P\overline{3}c1$ and $C2/m$ structures, the
$C2$ structure has 12 atoms in its primitive unit cell.

\section{Summary and Conclusions}

In summary, I have presented the phonon dispersions of the
$2\times2\times2$ $P\overline{3}c1$ phase of TiSe$_2$, which is the
currently accepted low-temperature structure of this material.  They
show weak instabilities in the acoustic branches, suggesting that this
phase might not be the ground state. To find the lowest energy
structure, I studied the energetics of all possible structures
corresponding to the isotropy subgroups due to the $M_1^-$ and $L_1^-$
phonon instabilities present in the parent $P\overline{3}m1$ phase.
The structure with the lowest energy has the space group $C2$ and
order parameter $M_1^- (a,0,0) + L_1^- (0,b,b)$.  The primitive unit
cell of this phase has 12 atoms, and it lacks inversion symmetry.

\section{acknowledgements}
This work was supported by Agence Nationale de la Recherche under
grant no.\ ANR-19-CE30-0004.  The computational resources were provided
by GENCI-CINES (grant A0090911099) and the Swiss National
Supercomputing Center (grant s820).

\newpage

\section{Supplemental Material}

% \section{Crystal structures}

\begin{table}[H]
  \caption{\label{tab:st164} Calculated atomic coordinates of
    TiSe$_2$ in the parent $P\overline{3}m1$ phase obtained using the
    optb88-vdw functional. Calculated lattice parameters are $a = b =
    3.55475$, $c = 6.080271$ \AA, $\alpha = \beta = 90^\circ$ and
    $\gamma = 120^\circ$. }
  \begin{ruledtabular}
    % \begin{tabular}{l l @{\hspace{1em}} d{1.5} d{1.5} d{1.5}}
    %   atom & site  & \multicolumn{1}{c}{$x$} & \multicolumn{1}{c}{$y$}
    %   & \multicolumn{1}{c}{$z$} \\
    \begin{tabular}{l l @{\hspace{1em}} c c c}
      atom & site  & $x$ & $y$ & $z$ \\
      \hline
       Ti    & $1a$ & 0 & 0 & 0  \\
       Se    & $2d$ & 1/3 & 2/3 & 0.25438 \\
    \end{tabular}
  \end{ruledtabular}
%  \vspace{0.5in}
\end{table}

\begin{table}[H]
  \caption{\label{tab:st165} Calculated atomic coordinates of TiSe$_2$
    in the $L_1^- (a,a,a)$ $P\overline{3}c1$ phase obtained using the
    optb88-vdw functional. Calculated lattice parameters are $a = b =
    7.11167$, $c = 12.17568$ \AA, $\alpha = \beta = 90^\circ$ and
    $\gamma = 120^\circ$. }
  \begin{ruledtabular}
    \begin{tabular}{l l @{\hspace{1em}} c c c}
      atom & site  & $x$ & $y$ & $z$ \\
      \hline
      Ti1    & $2a$ & 0 & 0 & 1/4  \\
      Ti2    & $6f$ & 0.50943 & 0 & 1/4 \\
      Se1    & $4d$ & 1/3 & 2/3 & 0.62316 \\
      Se2    & $12g$ & 0.66700 & 0.83055 & 0.87735 
    \end{tabular}
  \end{ruledtabular}
\end{table}

%\newpage

%\vspace{7in}
\begin{table}[H]
  \caption{\label{tab:st15} Calculated atomic coordinates of TiSe$_2$
    in the $L_1^- (a,a,b)$ $C2/m$ phase obtained using the optb88-vdw
    functional. Calculated lattice parameters are $a = 12.31768$, $b =
    7.11156$, $c = 12.17613$ \AA, $\alpha = 90^\circ$, $\beta =
    90.00156^\circ$ and $\gamma = 90^\circ$. }
  \begin{ruledtabular}
    \begin{tabular}{l l @{\hspace{1em}} c c c}
      atom & site  & $x$ & $y$ & $z$ \\
      \hline
      Ti1    & $4e$ &  0         & 0.00943 &  1/4 \\
      Ti2    & $4e$ &  0         & 0.50001 &  1/4 \\
      Ti3    & $8f$ &  0.25471   & 0.24528 &  0.25000 \\
      Se1    & $8f$ & $-0.08177$   & 0.24877 &  0.37735 \\
      Se2    & $8f$ &  0.66650   & 0.00296 &  0.37735 \\
      Se3    & $8f$ &  0.16667   & 0.00000 &  0.37684 \\
      Se4    & $8f$ &  0.41528   & 0.24827 &  0.37735 
    \end{tabular}
  \end{ruledtabular}
%  \vspace{0.9in}
\end{table}

\begin{table}[H]
  \caption{\label{tab:st5} Calculated atomic coordinates of TiSe$_2$
    in the $M_1^-(a,0,0) + L_1^- (0,b,b)$ $C2$ phase obtained using the optb88-vdw
    functional. Calculated lattice parameters are $a = 12.17894$, $b =
    7.11183$, $c = 8.66017$ \AA, $\alpha = 90^\circ$, $\beta =
    134.67258^\circ$ and $\gamma = 90^\circ$. }
  \begin{ruledtabular}
    \begin{tabular}{l l @{\hspace{1em}} c c c}
      atom & site  & $x$ & $y$ & $z$ \\
      \hline
      Ti1    &  $2b$  &  0   &  0.49045  &  1/2  \\
      Ti2    &  $2b$  &  0   &  0.00005  &  1/2  \\
      Ti3    &  $4c$  &  0.74521   &  0.25469  &  $-0.00960$ \\
      Se1    &  $4c$  &  0.79384   &  0.49698  &  0.83295  \\
      Se2    &  $4c$  &  0.54259   &  0.25178  &  0.33048  \\
      Se3    &  $4c$  &  0.79347   &  0.00000  &  0.83333  \\
      Se4    &  $4c$  &  0.04560   &  0.25121  &  0.33650  
    \end{tabular}
  \end{ruledtabular}
\end{table}

\end{document}